\documentstyle[12pt]{article}

                                 \newcommand{\sect}[1]{\setcounter{equation}{0}\section{#1}}
                                 \newcommand{\subsect}[1]{\subsection{#1}}

                                 \def\be{\begin{equation}}
                                 \def\ee{\end{equation}}
                                 \def\bea{\begin{eqnarray}}
                                 \def\eea{\end{eqnarray}}

                                 \def\1{\'{\i}}
                                 \def\R{{\mathbb R}}

                                 \def\o{\Omega}

                                 \def\>#1{{\bf #1}}

                                 \def\1{\'{\i}}                           
                                 \def\R{\rm I\kern-.2em R} 
                                 \def\C{\rm I\kern-.5em C} 
                                 \def\back{\!\!\!\!\!\!}

                                 \def\tfrac#1#2{ {\scriptstyle { \frac {#1}{#2}}}}         
                                          
                                 \def\conm#1#2{\left [ {#1},{#2} \right ]}

\begin{document}

                                 \title{
                                 Effective $su_q(2)$ models and polynomial algebras for fermion-boson
                                 Hamiltonians}

                                 \author{A. Ballesteros$^1$, O. Civitarese$^2$, F.J. Herranz$^1$ and M.
                                 Reboiro$^2$\\
                                 \\
                                                 \ $^1$ {\it  Departamento de F\1sica, Universidad de
                                 Burgos,}\\
                                                 {\it  Pza. Misael Ba\~nuelos s.n., E-09001-Burgos, Spain}
                                 \\
                                 \\
                                         \ $^2$ {\it  Departamento de F\1sica, Universidad Nacional de La
                                 Plata}\\
                                                 {\it  c.c.~67 1900, La Plata, Argentine}
                                         }

                                  \maketitle

                                 \begin{abstract}

                                 Schematic  
                                 $su(2)\oplus h_3$ interaction Hamiltonians, where $su(2)$ plays the role
                                 of the pseudo-spin algebra
                                 of fermion operators and $h_3$ is the Heisenberg algebra for bosons, are shown
                                 to be closely related to certain nonlinear models defined on a single quantum
                                 algebra
                                 $su_q(2)$ of quasifermions. In particular,
                                 $su_q(2)$ analogues of the Da Providencia-Sch\"utte and extended Lipkin models
                                 are presented. The connection between $q$ and the physical parameters of the
                                 fermion-boson system is analysed, and the integrability properties of the
                                 interaction Hamiltonians are discussed by using polynomial algebras.

                                 \end{abstract}


                                 \sect{Introduction} 

                                 The aim of this contribution is to discuss the
                                 equivalence between systems of interacting fermions and bosons and
                                 systems of $q$-deformed effective fermions.  The schematic fermion-boson
                                 interaction Hamiltonians that we
                                 shall deal with are of two different types:
                                 \begin{eqnarray}
                                 && H_{++}= \omega_f (T_0 +\Omega) +
                                           \omega _b B^{\dagger}B + G\,\left( T_{+}^k \, (B^{\dagger})^p\, 
                                 +\, T_{-}^k\,  (B)^p \right)
                                 \label{masmas}\\
                                 && H_{+-}=\omega_f (T_0 +\Omega) +
                                           \omega _b B^{\dagger}B + G\,\left( T_{+}^k \, (B)^p\, +\,
                                 T_{-}^k \, (B^{\dagger})^p \right)
                                 \label{masmenos}\end{eqnarray}
                                 and in this paper we shall restrict our study to the cases with $k,m=1,2$.
                                 The (two-level) fermions are represented by the collective pseudospin
                                 operators $\{T_0,T_\pm\}$ and the
                                 energy difference between fermion levels is fixed by the energy scale
                                 $\omega_f$.  Fermions are coupled to
                                 an external boson field with frequency $\omega_b$ that is quantized
                                 through the creation
                                 (annihilation) operators $B^{\dagger}$ ($B$). 

                                 Many interesting quantum models fall into this category of Hamiltonians.
                                 In particular, $H_{+-}$ with
                                 $k=p=1$ is the well-known Dicke model in Quantum Optics \cite{Dicke},
                                 that exhibits many of the
                                 characteristic features of quantum nonlinear  phenomena \cite{Koz}.
                                 Models based on
                                 the coupling between bi-fermions and bosons have been introduced long ago
                                 \cite{r7,r8} (see also the review paper \cite{r6} and references therein) 
                                 and are particularly
                                 suitable to describe condensation phenomena and transitions from
                                 fermionic to bosonic phases. Among
                                 them, we shall study the Da Providencia-Sch\"utte model 
                                 \cite{r4} (that corresponds to the case $H_{++}$ with $k=p=1$) and two
                                 different extensions ($k=2$ and
                                 $p=1$) \cite{r11} of the Lipkin-Meshkov-Glick model ($k=2$ and
                                 $p=0$) \cite{r11,r1}. 

                                 It turns out that all the abovementioned models can be replaced by
                                 effective $su_q(2)$
                                 quasifermion Hamiltonians with no boson operators, provided the
                                 deformation parameter $q$ is fitted in
                                 a suitable way in terms of physical constraints. 
                                 We  will refer to numerical analysis that strongly confirm this statement 
                                 \cite{BCHR} and we shall
                                 comment on some of the integrability properties of this class of
                                 interactions.

                                 In this respect, we recall that the
                                 exact solvability of the
                                 $su_q(2)$ interaction Hamiltonian
                                 \begin{equation}
                                 H_q^{\mbox{int}}= q^{  \frac {\tilde{T}_0}{2}}  ({\tilde{T}}_+
                                 +{\tilde{T}}_-) q^{\frac {\tilde{T}_0}{2}}
                                 \label{hqint}  
                                 \end{equation}
                                 was already found in \cite{BCh}, and this operator will be used as the
                                 building block for the effective
                                 $q$-Hamiltonians that we are going to introduce. Explicitly, the
                                 eigenvalues of (\ref{hqint})
                                 are just the $q$-numbers $[2\,m]_q$ ($m=-j,\dots,+j$), and the associated
                                 eigenvectors can be expressed in terms of
                                 $q$-Krawtchouk polynomials \cite{r24}. 
                                 On the other hand, the use of quantum deformations of $su(2)$ in this
                                 context is quite natural from
                                 a purely algebraic point of view, since it is well known that 
                                 fermion-boson models are related to polynomial generalizations of the
                                 $su(2)$ algebra (see, for instance,  \cite{Kar}). In particular, if we
                                 consider the operators
                                 \be
                                 K_0=T_0 \qquad\qquad K_+=T_{+}^k \, (B^{\dagger})^p\qquad\qquad  
                                 K_-=T_{-}^k\,  (B)^p
                                 \ee
                                 and we take a representation of them on a certain invariant subspace of
                                 the interaction
                                 Hamiltonian, we get a commutation rule of the type
                                 \be
                                 \conm{K_+}{K_-}=F(K_0)
                                 \ee
                                 where $F(K_0)$ is a polynomial of $K_0$ with degree $(2\,k+p-1)$, and the 
                                 same is true if we consider the
                                 operators
                                 $K_+=T_{+}^k \, (B)^p$ and $K_-=T_{-}^k\,  (B^{\dagger})^p$. 

                                 In this paper, we shall use this
                                 polynomial algebra approach to get new (to our knowledge) integrals of
                                 the motion for all the models
                                 under study. Moreover, since the coefficients of the polynomial $F(K_0)$
                                 are given in terms of the
                                 physical parameters of the model (degeneracy of the fermion shells,
                                 quantum numbers of the invariant
                                 subspace, etc.) we will be able to obtain new interpretations of the
                                 transition to certain dynamical
                                 regimes (for instance, the strong field limit \cite{Koz}) in terms of
                                 algebraic transformations like
                                 contraction processes
                                 \cite{IW}. In general, we expect that the comparison between the
                                 algebraic properties of such polynomial
                                 algebras and those of $su_q(2)$ will explain in more fundamental terms
                                 the efficiency of quantum
                                 algebras in order to model effective fermion-boson interactions.

                                 \sect{The Da Providencia-Sch\"utte model} 

                                 The model proposed
                                 by Da Providencia and Sch\"utte (DPS) is a solvable model which
                                 exhibits a phase transition between nucleonic and pionic
                                 condensates and consists of $N=2\,\Omega$  fermions moving
                                 in two single-shells each with degeneracy $2\,\Omega$. The DPS
                                 Hamiltonian reads \cite{r4}
                                 \begin{eqnarray}
                                 H_{  } = \omega_f (T_0 +\Omega) +
                                           \omega _b B^{\dagger}B
                                 + G (T_{+} B^{\dagger}+T_{-} B), \label{aa}
                                 \end{eqnarray}
                                 where $G$ is the strength of the interaction and $\{T_0,T_\pm\}$
                                 are the generators of the $su(2)$ algebra of collective fermions:
                                 \begin{equation}
                                 [T_0,T_+]=T_+,\quad [T_0,T_-]=-T_-,\quad [T_+,T_-]=2\,T_0 .
                                 \label{ac}
                                 \end{equation}

                                 The Hamiltonian (\ref{aa}) commutes with the operator
                                 \begin{equation}
                                 P =B^{\dagger} B-(T_0+\Omega).
                                 \label{ae}
                                 \end{equation}
                                 Therefore, the matrix elements of $H$ can be computed in a basis $|
                                 m_{\Omega},n \rangle$
                                 labeled by the eigenvalues of the number operators for 
                                 fermions and bosons.
                                 In this basis the eigenvalues of $P$ are given by
                                 \begin{equation}
                                 P | m_{\Omega},n \rangle = (n-m_{\Omega}-\Omega) | m_{\Omega},n \rangle.
                                 \label{ag}
                                 \end{equation}
                                 In particular, we shall diagonalize $H$ in the subspace spanned by
                                 the states $| m_{\Omega},L+m_{\Omega}+\Omega \rangle \equiv|
                                 m_{\Omega};L,\Omega \rangle $ which have a fixed eigenvalue $L$ of
                                 $P$
                                 \begin{equation}
                                 P | m_{\Omega};L,\Omega \rangle = L | m_{\Omega};L,\Omega \rangle.
                                 \label{ag2}
                                 \end{equation}
                                 In this subspace, the non-zero matrix elements of $H$ are \cite{BCHR}
                                 \bea
                                 && \back\back  \langle m_{\Omega};L,\Omega| H | m_{\Omega};L,\Omega \rangle =
                                 \omega_b L + (\omega_f+\omega_b) (\Omega+m_{\Omega}),  \label{9a} \\
                                 && \back\back  \langle m_{\Omega}+1;L,\Omega| H | m_{\Omega};L,\Omega
                                 \rangle  = \cr
                                 &&\qquad\qquad   G \sqrt { (\Omega
                                 +m_{\Omega}+1) (\Omega-m_{\Omega})(L+\Omega+m_{\Omega}+1)}. 
                                 \label{9b}
                                 \eea
                                 The dimension of such subspace depends on the sign of $L$.
                                 For $L \geq 0$
                                 the quantum number  $m_{\Omega}$ can take the values
                                 \begin{eqnarray}
                                  m_{\Omega}=-\Omega, -\Omega+1,..., \Omega,
                                 \label{ai}
                                 \end{eqnarray}
                                 and the subspace has dimension $2\Omega+1$.
                                 If $L<0$, $m_{\Omega}$ takes the values
                                 \begin{eqnarray}
                                 m_{\Omega}=-L-\Omega,- L-\Omega+1,..., \Omega, \label{aj}
                                 \end{eqnarray}
                                 and accordingly, the dimension of the invariant subspace is $2\,\Omega+L+1$.

                                 In general, we shall say that the system is in a
                                 {\it normal} phase when the correlated ground
                                 state is the eigenstate of the symmetry operator $P$ with the
                                 eigenvalue $L=0$. The denomination {\it deformed} phase will be
                                 assigned to cases where the correlated ground state is an
                                 eigenstate of $P$ with eigenvalue $L \neq 0$: if $L>0$ we shall have a
                                 so-called {\it bosonic} phase and
                                 the case $L<0$ corresponds to a {\it fermionic} one.

                                 \subsect{The DPS algebra}

                                 Let us define the operators
                                 \be
                                 K_0=T_0 \qquad\qquad K_+= T_{+} \, B^{\dagger}   \qquad\qquad K_-= T_{-} \, B.
                                 \ee
                                 For a given $\Omega$ and $L\geq 0$, if we consider the action of these
                                 operators within the subspace $|
                                 m_{\Omega};L,\Omega \rangle$ the following DPS algebra is obtained:
                                 \bea
                                 && \conm{K_0}{K_\pm}=\pm\,K_\pm\cr
                                 && \conm{K_+}{K_-}=-\o(\o + 1) + (1+2\,L+2\,\o)\,K_0 + 3\,K_0^2.
                                 \label{dps}
                                 \eea
                                 It can also be checked that the DPS algebra for the $L<0$ case is just
                                 (\ref{dps}) where $L$
                                 is replaced by $-L$.
                                 Therefore, a new integral of the motion for the DPS model is given by the 
                                 Casimir operator of the
                                 DPS-algebra, that can be found by standard methods
                                 \cite{Rocek} and reads
                                 \be
                                 C=K_+\,K_- + K_0^3 + (\,L+\o - 1)\,K_0^2 - \{(\,L+\o)+\o(\o + 1)\}\,K_0.
                                 \ee
                                 In the corresponding $L\geq 0$ subspace, the eigenvalue of $C$ is
                                 \be
                                 \o(\o + 1)(\o + L).
                                 \ee

                                 We remark that the DPS algebra (\ref{dps}) is a quadratic generalization
                                 of the $su(2)$ Lie algebra. In
                                 fact, the
                                 $su(2)$ algebra can be obtained as a contraction
                                 \cite{IW} of the DPS algebra in the
                                 $L\to\infty$ limit. Namely, if we define the ``contracted generators" as
                                 \be
                                 J_0=K_0 \qquad\qquad J_\pm=\frac{1}{\sqrt{L}} K_{\pm} 
                                 \label{cont}
                                 \ee
                                 and we compute their commutation rules we get
                                 \bea
                                 && \conm{J_0}{J_\pm}=\pm\,J_\pm\cr
                                 && \conm{J_+}{J_-}=\frac{1}{{L}}\{
                                 -\o(\o + 1) + (1+2\,L+2\,\o)\,J_0 + 3\,J_0^2\}
                                 \eea
                                 which is still isomorphic to the DPS algebra. However, the $L\to\infty$
                                 limit of these commutation rules
                                 gives
                                 \be
                                 \conm{J_0}{J_\pm}=\pm\,J_\pm \qquad\qquad
                                 \conm{J_+}{J_-}=2\,J_0. 
                                 \ee
                                 In physical terms, the $L\to\infty$ limit is just the well-known ``strong 
                                 field" limit of the
                                 Dicke model in Quantum Optics \cite{Koz}, for which the interaction
                                 dynamics is given by $su(2)$. If we
                                 rewrite the DPS Casimir in terms of the contracted generators we find
                                 \be
                                 C=L\,J_+\,J_- + J_0^3 + (\,L+\o - 1)\,J_0^2 - \{(\,L+\o)+\o(\o + 1)\}\,J_0
                                 \ee
                                 and the $su(2)$ Casimir operator $C_{su(2)}$ is obtained by computing
                                 \be
                                 C_{su(2)}=\lim_{L\to\infty}{\frac{C}{L}}=J_+\,J_- +J_0(J_0-1).
                                 \ee
                                 Note that the contracted eigenvalues are just $\o(\o + 1)$, as it should
                                 be. As we shall see in the
                                 sequel, this polynomial algebra approach can be applied to all the
                                 fermion-boson Hamiltonians under
                                 consideration.

                                 \subsect{Effective $su_q(2)$ Hamiltonians for the DPS model}

                                 The quantum algebra $su_q(2)$ is a Hopf algebra
                                 deformation of $su(2)$ \cite{r23} with generators
                                 $\{{\tilde T}_\pm,{\tilde T}_0\}$  and commutation rules
                                 \begin{equation}
                                 [{\tilde T}_0,{\tilde T}_\pm]=\pm {\tilde T}_\pm,\quad  [{\tilde
                                 T}_+,{\tilde T}_-]=[2\,
                                 {\tilde T}_0]_q,
                                 \label{ak}
                                 \end{equation}
                                 where the $q$-number $[x]_q$ is defined by
                                 \begin{equation}
                                 [x]_q  =   \frac{q^x-q^{-x}}{q-q^{-1}}    =   \frac {\sinh (z \;
                                 x)}{\sinh (z)}.
                                 \label{al}
                                 \end{equation}
                                 We shall use alternatively $q$ and $z$ (where
                                 $q={\rm e}^z$) as the deformation parameter, and we shall
                                 assume that $q$ is real. The $su(2)$ algebra
                                 (\ref{ac}) is recovered from (\ref{ak}) in the limit $q\to 1$
                                 ($z\to 0$).

                                 When $q$ is not a root of unity, the irreducible representations
                                 of $su_q(2)$ are obtained as a straightforward generalization of
                                 those of $su(2)$ \cite{CP,GS}:
                                 \begin{eqnarray}
                                 && {\tilde{T}}_0 | j , m \rangle =m\,| j , m \rangle , \nonumber  \\
                                 && {\tilde{T}}_+ | j , m \rangle =\sqrt{ [j+m+1]_q [j-m ]_q}\,| j , m+1
                                 \rangle ,
                                 \nonumber \\ && {\tilde{T}}_- | j , m \rangle =\sqrt{ [j-m+1]_q [j+m
                                 ]_q}\,| j , m-1
                                 \rangle .
                                 \label{am}
                                 \end{eqnarray}

                                 By following \cite{BCHR,BCh}, we consider an effective
                                 Hamiltonian defined as
                                 \begin{equation}
                                 H_q = \omega_b L + (\omega_b + \omega_f ) ({\tilde{T}}_0 + \Omega)
                                 + \chi(q)  q^{  \frac {\tilde{T}_0}{2}}  ({\tilde{T}}_+
                                 +{\tilde{T}}_-) q^{\frac {\tilde{T}_0}{2}} \label{an}
                                 \end{equation}
                                 where $\chi(q)$ is a scalar function to be fixed, and $H_q$ will be
                                 realized in
                                 a $su_q(2)$ irreducible representation with the same dimension as
                                 the subspace spanned by $| m_{\Omega};L,\Omega \rangle $
                                 (therefore, with $j=j(\Omega, L)$). The
                                 non-vanishing matrix elements of (\ref{an}) read
                                 \bea
                                 &&\langle j , m |  H_q | j , m \rangle  = \omega_b L  +
                                 (\omega_f+\omega_b) (m+\Omega) ,
                                  \label{16a} \\
                                 && \langle j , m+1 |  H_q | j , m \rangle  =
                                 \chi(q) {q}^{(m+\frac12)}\sqrt{ [j+m+1]_q [j-m ]_q}. \label{16b}
                                 \eea
                                 In order to fit the dimension of the representation with respect to the
                                 invariant subspace of the
                                 DPS model, we have to take
                                 $j=\Omega$ and $m=m_{\Omega}$   for the effective $L \ge 0$ model,
                                 while for $L<0$, $j=\Omega+ \frac L2$ and $m=m_{\Omega}+\frac L2$.

                                 The main conclusion of \cite{BCHR} (see also \cite{BCh} for the Dicke
                                 model) is that the Hamiltonian
                                 (\ref{aa}) is essentially equivalent to (\ref{an}). In other words, the
                                 bosonic degrees of freedom
                                 included in (\ref{aa}) may be reabsorbed by the $q$-deformation in
                                 (\ref{an}) provided that
                                 $q$ is defined as an appropriate function of both $\Omega$ and
                                 $L$. In this way it is possible to regard $H_q$ as an
                                 effective Hamiltonian with physical properties similar to those of
                                 $H$. As it is extensively shown in \cite{BCHR} through numerical studies, 
                                 both the ground state energy and the full spectrum of the DPS model 
                                 can  accurately be reproduced by using the effective $q$-Hamiltonian
                                 (\ref{an}).

                                 \subsection{The q-DPS algebra}

                                 By following the same algebraic approach leading to the previous DPS
                                 algebra, now we should consider the
                                 $su_q(2)$ operators
                                 \be
                                 K_0=\tilde{T}_0 \qquad\qquad
                                 K_+= q^{\frac {\tilde{T}_0}{2}} \,{\tilde{T}}_+
                                 \, q^{\frac {\tilde{T}_0}{2}}   \qquad\qquad 
                                 K_-= q^{\frac {\tilde{T}_0}{2}} \,{\tilde{T}}_-
                                 \, q^{\frac {\tilde{T}_0}{2}}
                                 \ee
                                 such that the effective Hamiltonian (\ref{an}) is a linear function of
                                 $K_0$ and $K_\pm$.
                                 In this new basis, the $q$-commutation rules of $su_q(2)$  read 
                                 \be
                                 [K_0,K_\pm]=\pm\,K_\pm,\qquad\qquad 
                                 q\,K_+\,K_- - q^{-1}\,K_-\,K_+ =q^{2\,K_0}\,[2\,K_0]_q
                                 \label{newconm}
                                 \ee
                                 and these expressions hold for any irreducible representation $j$ of
                                 $su_q(2)$.
                                 The Casimir element
                                 for this algebra
                                 is
                                 \be
                                 C_q=[K_0]_q\,[K_0-1]_q + q^{-2\,K_0+1}\,K_+\,K_-,
                                 \label{ncas}
                                 \ee
                                 and its eigenvalue is just $[j]_q\,[j+1]_q$. Obviously, $C_q$ is an
                                 integral of the motion for $H_q$
                                 (\ref{an}). By working on a fixed irreducible representation $j$, the
                                 latter $q$-commutator can be
                                 rewritten as the following commutation rule (that hereafter we will call
                                 the $q$-DPS algebra):
                                 \be
                                 \conm{K_+}{K_-}=\frac{q^{2j+2}+q^{-2j}}{1-q^{2}}\,q^{2\,K_0} - 
                                 \frac{1+q^{2}}{1-q^{2}}\,q^{4\,K_0},
                                 \label{qconm}
                                 \ee
                                 which should have algebraic properties closely related to the ones of the 
                                 DPS algebra (\ref{dps}), since
                                 both models are physically equivalent. In fact, the analytic fitting
                                 $q=q(\Omega,L)$ should be found by comparing the properties of the DPS
                                 and $q$-DPS algebras. Work on
                                 this open problem is actually in progress \cite{BCHRp}.

                                 \sect{Extended Lipkin models}

                                 As a second example of Hamiltonians including fermionic and
                                 bosonic degrees of freedom, let us introduce the extended
                                 Lipkin-Meshkov-Glick Hamiltonian (LE
                                 model) \cite{r11}, which is just (\ref{masmenos}) with $k=2$ and $p=1$:
                                 \begin{eqnarray}
                                 H_{+-} =  \omega_f (T_0 +\Omega) +
                                           \omega_b B^{\dagger}B
                                 + G (T_{+}^2 B+T_{-}^2 B^{\dagger}).
                                 \label{b1a}
                                 \end{eqnarray}
                                 which commutes
                                 with the operator
                                 \cite{twobosons}
                                 \begin{equation}
                                 P_{(+)}  = B^{\dagger} B+ \frac {1}{2} (T_0 +\Omega) .
                                 \label{b1b}
                                 \end{equation}
                                 Therefore, the matrix elements of the LE model $H_{+-}$ can be calculated 
                                 in a basis
                                 labeled by the eigenvalues of $P_{(+)}$:
                                 \begin{equation}
                                 P_{(+)}  | m_{\Omega},n \rangle = L | m_{\Omega}, n \rangle=
                                 (n+ \frac 12 (\Omega+m_{\Omega})) | m_{\Omega}, n \rangle,
                                 \label{b1c}
                                 \end{equation}
                                 and we shall consider the invariant subspace  with $L$
                                 fixed:
                                 \be
                                 | m_{\Omega},L-\frac 12 (\Omega+m_{\Omega}) \rangle \equiv|
                                 m_{\Omega};L,\Omega \rangle .
                                 \ee
                                 In this subspace the non-zero matrix elements of $H$ read
                                 \bea
                                 && \back\back \langle m_{\Omega};L,\Omega| H_{+-} | m_{\Omega};L,\Omega
                                 \rangle =
                                 \omega_b L + (\omega_f -\frac 12 \omega_b)(\Omega+m_{\Omega}), \\
                                 && \back\back \langle m_{\Omega}+2;L,\Omega| H_{+-} | m_{\Omega};L,\Omega 
                                 \rangle =
                                 G \sqrt {L- \frac 12 (\Omega+m_{\Omega})} \times  \nonumber\\
                                 && \qquad\sqrt {(\Omega+m_{\Omega}+2) (\Omega+m_{\Omega}+1)
                                 (\Omega-m_{\Omega})
                                 (\Omega-m_{\Omega}-1) },
                                 \label{b1d}
                                 \eea
                                 and we have to distinguish the following classes of invariant subspaces:
                                 \begin{eqnarray}
                                 && L \ge \Omega, L \; {\rm integer}, \qquad
                                 m_{\Omega}+\Omega=0,2,..., 2\Omega,
                                 \nonumber \\
                                 && L > \Omega, L \; {\rm half\ integer}, \qquad
                                 m_{\Omega}+\Omega=1,3,..., 2\Omega-1,
                                 \nonumber \\
                                 && L < \Omega, L \; {\rm integer} , \quad m_{\Omega}+\Omega=
                                 0,2,...,2 L,
                                 \label{b1e} \\
                                 && L < \Omega, L \; {\rm half\ integer} ,\qquad m_{\Omega}+\Omega=
                                 0,2,...,2L-1. \nonumber
                                 \end{eqnarray}

                                 Another Lipkin-type Hamiltonian can also be defined through
                                 (\ref{masmas}) with $k=2$ and $p=1$:
                                 \begin{eqnarray} 
                                 H_{++} = \omega_f (T_0 +\Omega) +
                                           \omega_b B^{\dagger}B
                                 + G (T_{+}^2 B^{\dagger}+T_{-}^2 B),
                                 \label{b2a}
                                 \end{eqnarray}
                                 which differs from
                                 (\ref{b1a}) in the ground state correlations \cite{r11}. Since $H_{++}$
                                 commutes with the operator
                                 \begin{equation}
                                 P_{(-)} = B^{\dagger} B- \frac {1}{2} (T_0 +\Omega),
                                 \label{b2b}
                                 \end{equation}
                                 its matrix elements can be calculated in a basis labeled by the
                                 eigenvalues of $P_{(-)}$. Namely
                                 \begin{equation}
                                 P_{(-)}  | m_{\Omega},n \rangle = (n- \frac 12
                                 (\Omega+m_{\Omega})) | m_{\Omega},n \rangle, \label{b1c1}
                                 \end{equation}
                                 and in this case we shall compute the matrix elements within the subspace 
                                 spanned by
                                 the states
                                 $| m_{\Omega},L+\frac 12 (\Omega+m_{\Omega}) \rangle \equiv|
                                 m_{\Omega};L,\Omega \rangle $.
                                 The non-zero matrix elements of $H_{++}$ (\ref{b2a})  are now given by
                                 \bea
                                 &&\back\back \langle m_{\Omega};L,\Omega| H_{++} | m_{\Omega};L,\Omega \rangle
                                 = \omega_b L + (\omega_f+ \frac 12 \omega_b)(\Omega +m_{\Omega}) , \\
                                 &&\back\back \langle m_{\Omega}+2;L,\Omega| H_{++} | m_{\Omega};L,\Omega
                                 \rangle  =
                                 G \sqrt {L+ \frac 12 (\Omega+m_{\Omega})+1} \times \nonumber\\
                                 && \qquad \sqrt { (\Omega+m_{\Omega}+2) (\Omega+m_{\Omega}+1)
                                 (\Omega-m_{\Omega}) (\Omega-m_{\Omega}-1)
                                 }, 
                                 \label{b2d}
                                 \eea
                                 where the dimension of the subspace depends on $L$ and $\Omega$, since we 
                                 have to consider the following
                                 possibilities for the set of values of the quantum number $m_{\Omega}$:
                                 \begin{eqnarray}
                                 && L \ge 0, L \; {\rm integer}, \quad m_{\Omega}+\Omega=0,2,..., 2\Omega, 
                                 \nonumber \\
                                 && L > 0, L \; {\rm half\ integer}, \qquad
                                  m_{\Omega}+\Omega=1,3,..., 2\Omega-1, \label{b2e}\\
                                 && L < 0, L \; {\rm integer} ,\quad   m_{\Omega}+\Omega=-2 L ,-2
                                 L+2,..., 2\Omega, \nonumber
                                 \\ && L < 0, L \; {\rm half\ integer} , \qquad
                                 m_{\Omega}+\Omega=-2 L,-2 L+2,...,2\Omega-1. \nonumber
                                 \end{eqnarray} 

                                 \subsect{LE algebras and their Casimir operators}

                                 The polynomial algebra approach can also be used for the LE models
                                 (\ref{b1a}) and (\ref{b2a}). We start
                                 our analysis by recalling the cubic algebra linked to the original
                                 Lipkin-Meshkov-Glick (LMG)
                                 Hamiltonian \cite{r1}, since the latter will appear as the strong field
                                 limit for the  extended
                                 LE models.

                                 \subsubsection{The  LMG algebra}
                                 We recall the $su(2)$ LMG Hamiltonian \cite{r1} given by
                                 \begin{equation}
                                 H= \omega_f\, {T}_0 + \chi\,({ {T}}_+^2
                                 +{ {T}}_-^2).
                                 \end{equation}
                                 If we define (see also \cite{NLMG})
                                 \be
                                 K_0=T_0 \qquad\qquad 
                                 K_+= T_{+}^2    \qquad\qquad K_-= T_{-}^2 
                                 \ee
                                 we get the cubic algebra
                                 \bea
                                 && \conm{K_0}{K_\pm}=\pm\,2\,K_\pm\cr
                                 && \conm{K_+}{K_-}= 4\{2\,\o(\o + 1)-1\}\,K_0 -8\,K_0^3
                                 \label{lip}
                                 \eea
                                 where we have identified $j\equiv\o\geq0$. Note that this algebra is
                                 isomorphic to the Higgs algebra
                                 \cite{Higgs,Zhedanov} for any value of
                                 $\o$. The Casimir operator for this algebra is found to be:
                                 \be
                                 C=K_+\,K_- - K_0^4 + 4\, K_0^3 + \{2\o(\o+1)-5\}\,K_0^2 -
                                 2\{2\o(\o+1)-1\}\,K_0
                                 \label{cashiggs}
                                 \ee
                                 and the eigenvalues of this operator are
                                 $
                                 \o(\o^2 - 1)(\o +2).
                                 $

                                 \subsubsection{The LE$_{+-}$ algebra}

                                 From (\ref{b1a}), we can consider the operators
                                 \be
                                 K_0=T_0 \qquad\qquad 
                                 K_+= T_{+}^2 \, B   \qquad\qquad K_-= T_{-}^2 \, B^{\dagger}.
                                 \label{lepm}
                                 \ee
                                 If we compute its action on an invariant subspace of the type $L \ge \Omega$, 
                                 we obtain the following quartic generalization of the $su(2)$ Lie algebra:
                                 \bea
                                 && \conm{K_0}{K_\pm}=\pm\,2\,K_\pm\cr
                                 && \conm{K_+}{K_-}=\alpha_0+\alpha_1\,K_0
                                 +\alpha_2\,K_0^2+\alpha_3\,K_0^3+\alpha_4\,K_0^4
                                 \label{pm}
                                 \eea
                                 whose structure constants $\alpha_i(\o,L)$ read
                                 \bea
                                 && \alpha_0= \o(\o^2-1)(\o+2)   \cr
                                 && \alpha_1= 2(1-2\o(\o+1))(\o-2L-1)    \cr
                                 && \alpha_2= 7-6 \o(\o+1) \label{sc}\\
                                 && \alpha_3= 4(\o-2L-1)   \cr
                                 && \alpha_4= 5  . 
                                 \nonumber
                                 \eea
                                 The Casimir operator for (\ref{lepm}) can also be computed and gives a
                                 new integral of the motion for
                                 $H_{+-}$:
                                 \bea
                                 && \back\back C=K_+\,K_- + \beta_0+ \beta_1\,K_0
                                 +\beta_2\,K_0^2+\beta_3\,K_0^3+\beta_4\,K_0^4+\beta_5\,K_0^5
                                 \label{ss}\\
                                 && \mbox{with}\quad  \beta_0= -\o(\o^2 - 1)(\o +2)    \cr
                                 && \phantom{with}\quad \beta_1= 2(L+1) + \tfrac{1}{2}\o(-12
                                 +\o(\o^2+6\o-5)-8(\o+1)L )    \cr
                                 && \phantom{with}\quad\beta_2= -6-5 L + \tfrac{1}{2}\o(13
                                 +2\o(3-\o)+4(\o+1)L )  \cr
                                 && \phantom{with}\quad\beta_3= \tfrac{13}{2}-\o(\o+3) +4L  \cr
                                 && \phantom{with}\quad\beta_4= \tfrac{1}{2}\o-3-L   \cr
                                 && \phantom{with}\quad\beta_5= \tfrac{1}{2} 
                                 \nonumber  
                                 \eea
                                 The eigenvalue of $C$ is found to be
                                 $
                                 -\tfrac{1}{2}\o(\o^2 - 1)(\o +2)(\o-2L).
                                 $
                                 It can also be checked that the transformation (\ref{cont}) leads to the
                                 LMG algebra
                                 (\ref{lip}) and Casimir (\ref{cashiggs}) as the ``strong field" contraction
                                 $L\to\infty$  of (\ref{pm}) and (\ref{ss}), respectively. Similar quartic 
                                 algebras can be obtained for
                                 the remaining invariant subspaces \cite{BCHRp}.

                                 \subsubsection{The LE$_{++}$ algebra}

                                 For the second class of LE models (\ref{b2a}) we define the generators
                                 \be
                                 K_0=T_0 \qquad\qquad
                                 K_+= T_{+}^2 \, B^{\dagger}   \qquad\qquad K_-= T_{-}^2 \, B.
                                 \ee
                                 The associated quartic algebra for $L\geq 0$ reads:
                                 \bea
                                 && \back\back \conm{K_0}{K_\pm}=\pm\,2\,K_\pm\cr
                                 && \back\back \conm{K_+}{K_-}=\alpha_0+\alpha_1\,K_0
                                 +\alpha_2\,K_0^2+\alpha_3\,K_0^3+\alpha_4\,K_0^4
                                 \label{sc}\\
                                 && \mbox{with}\quad \alpha_0= -\o(\o^2-1)(\o+2)   \cr
                                 && \phantom{with}\quad \alpha_1= -2(1-2\o(\o+1))(\o+2L+1)    \cr
                                 && \phantom{with}\quad \alpha_2= -7+6 \o(\o+1) \cr
                                 && \phantom{with}\quad \alpha_3= -4(\o+2L+1)   \cr
                                 && \phantom{with}\quad \alpha_4= -5   
                                 \nonumber
                                 \eea
                                 The Casimir operator is now
                                 \bea
                                 && \back C=K_+\,K_- + \beta_0+ \beta_1\,K_0
                                 +\beta_2\,K_0^2+\beta_3\,K_0^3+\beta_4\,K_0^4+\beta_5\,K_0^5
                                 \label{scc}\\ && \mbox{with}\quad  \beta_0=
                                 \o(\o^2 - 1)(\o +2)    \cr && \phantom{with}\quad \beta_1= 2L -
                                 \tfrac{1}{2}\o(-4
                                 +\o(\o^2+6\o+3)+8(\o+1)L )    \cr && \phantom{with}\quad \beta_2= 1-5 L + 
                                 \tfrac{1}{2}\o(-9
                                 +2\o(\o-1)+4(\o+1)L )  \cr && \phantom{with}\quad \beta_3=
                                 -\tfrac{5}{2}+\o(\o+3) +4L  \cr
                                 && \phantom{with}\quad \beta_4= -\tfrac{1}{2}\o+2-L   \cr
                                 && \phantom{with}\quad \beta_5= -\tfrac{1}{2}   
                                 \nonumber
                                 \eea
                                 and the eigenvalue of $C$ is 
                                 $
                                 \tfrac{1}{2}\o(\o^2 - 1)(\o +2)(2+\o+2L).
                                 $
                                 Once more, the ``strong field" contraction of this algebra gives rise to
                                 the same LMG
                                 algebra (\ref{lip}), that underlies the asymptotic $L\to\infty$ dynamics
                                 of both LE models.

                                 \subsect{Effective $su_q(2)$ Hamiltonians for the LE models}

                                 Like in the case of the DPS  model,  an effective
                                 $q$-Hamiltonian for (\ref{b1a}) has been introduced in \cite{BCHR}
                                 \begin{equation}
                                 H_q = \omega_b L + ( \omega_f - \frac 12 \omega_b ) ( {\tilde{T}}_0 + \Omega )
                                 + \chi(q) q^{{\tilde{T}}_0}  ({\tilde{T}}_+^2 +{\tilde{T}}_-^2)
                                 q^{{\tilde{T}}_0},
                                 \label{ca}
                                 \end{equation}
                                 which has the following non-vanishing matrix elements
                                 \bea
                                 && \langle j , m |  H_q | j , m \rangle   =
                                 \omega_b L + (\omega_f -\frac 12 \omega_b) (\Omega+m), \label{cca} \\
                                 && \langle j , m+2 |  H_q | j , m \rangle = \chi(q) q^{2 (m+1)} \times  \cr
                                 &&\qquad\qquad  \sqrt { [j+m+2]_q [j+m+1]_q [j-m]_q [j-m-1]_q }. 
                                 \label{cb}
                                 \eea
                                 As the first step in order to fit the dimensions of
                                 (\ref{b1a}) and of (\ref{ca}) the appropriate
                                 relation $j=j(\Omega,L)$ has to be found and, as a consequence,
                                 $m=m(m_\Omega,\Omega,L)$ \cite{BCHR}. Afterwards, we have to consider as
                                 the effective
                                 Hamiltonian the restriction of the matrix elements
                                 (\ref{cca}) and (\ref{cb}) to the invariant subspace spanned by $| j , m
                                 \rangle$
                                 with $m=-j, -j+2,\dots, j-2, j$. In this way we obtain the
                                 effective matrix elements
                                 \begin{equation}
                                 \langle j , m+2 |  H_q | j , m \rangle  = \chi(q) \,h(L, \Omega, m_\Omega).
                                 \end{equation}
                                 For values of $L\ge \Omega$ ($L$ integer), we find $j=\Omega$,
                                 $m=m_\Omega$ and the
                                 function $h(L, \Omega, m_\Omega)$ is (the function $\chi(q)$ is also
                                 explicitly defined in \cite{BCHR}):
                                 \be
                                 h(L, \Omega, m_\Omega)= q^{2(m_\Omega+1)}
                                 \sqrt { [ {\Omega}+m_\Omega+2 ]_q [ {\Omega}+m_\Omega+1 ]_q [ {\Omega}-
                                 m_\Omega ]_q 
                                 [{\Omega}-m_\Omega-1 ]_q  }.
                                 \label{cc}
                                 \ee

                                 Afterwards, we have followed the same procedure as in the $q$-DPS model,
                                 and we have looked for
                                 values of the $q$ parameter (and, consequently, of $\chi(q)$) which may
                                 absorb bosonic degrees of
                                 freedom of (\ref{b1a}) and give rise to a similar spectrum for the purely 
                                 fermionic $q$-deformed
                                 Hamiltonian (\ref{ca}). The second type of LE model (\ref{b2a}) can also
                                 be approximated by the same
                                 type of
                                 $q$-Hamiltonian (\ref{ca}) and, in both cases,  numerical computations
                                 lead to an excellent fitting
                                 between the LE models and the effective $q$-Hamiltonians \cite{BCHR}.  
                                 Therefore,
                                 we conclude that certain interactions between fermions and bosons can
                                 accurately be  described by
                                 using
                                 $q$-fermions as quasiparticles (i.e; effective fermionic degrees
                                 of freedom) under the exactly solvable interaction given by the
                                 Hamiltonian $H_q^{\mbox{int}}$ (\ref{hqint}).

                                 From the mathematical point of view, a very interesting question to be
                                 solved is to find a suitable
                                 analytical expression of the deformation parameter $q$ in terms of the
                                 representation space labels $\o$
                                 and $L$. In this respect, the quartic LE algebras previously introduced
                                 should be relevant, since the
                                 $su_q(2)$ model (\ref{ca}) leads to the following natural definition of
                                 the $K$ operators
                                 \be
                                 K_0=\tilde{T}_0 \qquad\qquad
                                 K_+= q^{{\tilde{T}_0}} \,{\tilde{T}}_+^2
                                 \, q^{{\tilde{T}_0}}   \qquad\qquad 
                                 K_-= q^{{\tilde{T}_0}} \,{\tilde{T}}_-^2
                                 \, q^{{\tilde{T}_0}}
                                 \ee
                                 and their commutation rule $\conm{K_+}{K_-}$ (which is a generalization
                                 of (\ref{qconm})) has to carry
                                 essentially the same algebraic information as (\ref{pm}) and (\ref{sc}).
                                 Another intteresting feature
                                 appears in the analysis of the effective $q$-Hamil\-tonian for $H_{++}$
                                 (\ref{b2a}), since a $su(2)$
                                 symmetry of the model can dynamically be  restored for certain negative
                                 values of $L$ \cite{BCHR}. 
                                 A complete study of the LE algebras together with the abovementioned
                                 algebraic properties of these
                                 fermion-boson interactions will be addressed in a forthcoming paper
                                 \cite{BCHRp}.

                                 \bigskip
                                 \bigskip

                                 \noindent
                                 {\Large{{\bf Acknowledgments}}}

                                 \medskip

                                 \noindent This work has been partially supported by MCyT (Spain) under
                                 Project BFM2000-1055 and by
                                 CONICET (Argentine). M.R. acknowledge financial support of the Fundacion
                                 Antorchas and of Universidad de
                                 Burgos (Invited Professors Program).


\begin{thebibliography}{99}


                                 \bibitem{Dicke} R.H.\ Dicke, {Phys. Rev.}
                                 {\bf 93}, 99 (1954);
                                 G.\  Drobn\'y  and I.\ Jex, 
                                  Phys.\ Rev.\ A {\bf 46}, 500 (1992).

                                 \bibitem{Koz}  S.M.\ Chumakov and M.\ Kozierowski, Quant.\ 
                                 Semiclas.\  
                                   Optics {\bf 8}, 775 (1996).
                                  

                                 \bibitem{r7}{R. Eder, O. Rogojanu and G. A. Sawatzky,
                                 Phys. Rev. {\bf B 58} 7599 (1998);
                                  W. Hanke, R. Eder and E.
                                 Arrigoni, Physikalische Bl\"atter {\bf 54} 436 (1998).}

                                 \bibitem{r8}{B. Buck and C. V. Sukumar, Phys. Lett. {\bf 81A} 132 (1981); 
                                 ibid. J. Phys. {\bf A 17} 877 (1984);
                                  M. Tavis and F. W. Cummings,
                                 Phys. Rev. {\bf 170} 360 (1968); K. Hepp and E. H. Lieb, Ann.
                                 Phys. (N.Y.) {\bf 76} 360  (1973).}




                                 \bibitem{r6}{A. Klein and E.R. Marshalek, Rev. Mod. Phys. {\bf{63}}, 375
                                 (1991).}

                                 \bibitem{r4}{D. Sch\"utte and J. Da Providencia, Nucl. Phys. A {\bf
                                 {282}}, 518 (1977).}

                                 \bibitem{r11} {
                                 S. Jesgarz, S. Lerma H., P. O. Hess, O. Civitarese and M. Reboiro.
                                 Procceedings of the XXV Symposium of Nuclear Physics, Taxco 2002.
                                 M\'exico. Rev. Mex. Fis. (in press).}



                                 \bibitem{r1}{H. J. Lipkin, N. Meshkov and A. J. Glick, Nucl. Phys.
                                 {\bf{62}}, 188 (1965).}


                                 \bibitem{BCHR}{A. Ballesteros, O. Civitarese, F.J. Herranz and M. Reboiro,
                                 Phys. Rev. {\bf C 66}, 064317 (2002).}

                                 \bibitem{BCh}
                                   A.\ Ballesteros and S.M.\ Chumakov, J.\ Phys. A: \ Math. Gen.\ {\bf
                                 32}, 6261
                                 (1999).

                                 \bibitem{r24}{N. M. Atakishiyev and P. Winternitz, J.
                                 Phys. A:  Math. Gen. {\bf {33}}, 5303 (2000)}; 
                                  N. M. Atakishiyev and A.U. Klymyk, J.
                                 Phys. A:  Math. Gen. {\bf {35}}, 5267 (2002).

                                 \bibitem{Kar} V.P.\ Karassiov and A.B.\ Klimov, Phys.\ Lett.\
                                 A {\bf 189}, 43  (1994);  
                                  V.P.\ Karassiov, J.\ Sov.\ Laser Research  {\bf 13}, 188  (1992); ibid.
                                  {Phys.\ Lett.\ A}
                                   {\bf 238}, 19 (1998).


                                 \bibitem{IW} E. In\"on\"u,   and  E. P. Wigner,  {\it Proc. Natl. Acad.
                                 Sci. U. S.}  {\bf 39}, 510 (1953).

                                 \bibitem{Rocek}
                                   M. Rocek, Phys. Lett. {\bf B255}, 554
                                 (1991).
                                   
                                 \bibitem{r23}{P.P. Kulish and N. Reshetikhin, Zap. Nauch. Sem. LOMI{
                                 \bf{101}}, 101 (1981).}

                                 \bibitem{CP}{V. Chari and A. Pressley, {\it Quantum
                                 Groups} (Cambridge University Press, Cambridge, MA, 1994)}

                                 \bibitem{GS}{C. G\'omez, G. Sierra and  M. Ruiz-Altaba, {\it Quantum
                                 Groups in Two-Dimensional Physics} (Cambridge University Press,
                                 Cambridge, MA, 1996)}

                                 \bibitem{twobosons}{O. Civitarese and M. Reboiro, Phys. Rev. {\bf C 58},
                                 2787 (1998).}

                                 \bibitem{NLMG}
                                   N. Debergh and F.L. Stancu, J.\ Phys. A: \ Math. Gen.\ {\bf 34}, 3265
                                 (2001).

                                 \bibitem{Higgs}
                                   P.W. Higgs, J.\ Phys. A: \ Math. Gen.\ {\bf 12}, 309
                                 (1979).

                                 \bibitem{Zhedanov}
                                   A.S. Zhedanov, Mod. Phys. Lett. A. {\bf 7}, 507
                                 (1992).

                                 \bibitem{BCHRp}{A. Ballesteros, O. Civitarese, F.J. Herranz and M. Reboiro,
                                  in preparation.}


                                 \end{thebibliography}
                                 \end{document}